\documentstyle[preprint,aps]{revtex}
\begin{document}

\draft

\preprint{BUTP--95/18}

\title{Fixed point action and topology in the ${\rm CP}^{3}$ model}

\author{Rudolf Burkhalter}
\address{Institut f\"ur theoretische Physik, 
Universit\"at Bern,\\
Sidlerstrasse 5, CH--3012 Bern, Switzerland.}

\date{December 1995}

\maketitle

\begin{abstract}
We define a fixed point action in two-dimensional lattice ${\rm CP}^{N-1}$
models.  The fixed point action is a classical perfect lattice action, which
is expected to show strongly reduced cutoff effects in numerical simulations.
Furthermore, the action has scale-invariant instanton solutions, which enables
us to define a correct topological charge without topological defects.  Using
a parametrization of the fixed point action for the ${\rm CP}^{3}$ model in a
Monte Carlo simulation, we study the topological susceptibility.

\end{abstract}

\pacs{11.15.Ha 75.10.Hk 11.10.Hi}

\section{\bf Introduction}
\label{introsection}

One possible way of regularizing a continuum quantum field theory is to
introduce a lattice as a UV regulator. An additional benefit of this method is
that it opens the door to computer simulations. However, by naively
discretizing physical observables, the correct values are only obtained in the
continuum limit, when the lattice spacing is going to zero. At finite lattice
spacing the lattice induces systematic errors (cutoff effects). In order to
remove these cutoff effects one has to introduce finer and finer lattices, and
finally extrapolate the calculated quantities to the continuum limit.  This
delicate limit is a major difficulty in extracting continuum physics from the
lattice.

It is possible to circumvent this problem by using Wilson's renormalization
group theory. If one constructs the lattice action and all operators in a form
that corresponds to the renormalized trajectory, one obtains results that do
not depend on the lattice spacing. It has been shown in a series of papers
recently published \cite{HASENFRATZ,DEGRAND} that it is possible to construct
a fixed point action for asymptotically free theories. This action is the
fixed point (FP) of an exact renormalization group (RG) transformation and as
such may be taken as a first approximation to the renormalized trajectory.

As a pilot project a local parametrization of the FP action was constructed in
the O(3) nonlinear $\sigma$ model, and used in numerical simulations
\cite{HASENFRATZ}. The result was very promising: Although the FP action is
perfect strictly only classically, no cutoff effects were seen even at small
correlation lengths ($\xi \sim 3$).

In a subsequent paper \cite{BLATTER} a FP topological charge was proposed. It
was shown, that the combination FP action and FP charge has no topological
defects. The FP action has the correct value for instanton solutions, and does
not depend on the scale of the instanton. Hence it admits stable instanton
solutions on the lattice. This is in contrast to the standard lattice action,
whose value depends on the instanton scale, and which suffers from dislocations
\cite{LUSCHER}. These dislocations were suspected to be responsible for the
non scaling behavior of the topological susceptibility in the O(3) nonlinear
$\sigma$ model. However, using parametrizations of the FP action and the FP
topological charge in a Monte Carlo simulation, even with the absence of
dislocations a strong violation of scaling of the topological susceptibility
was established \cite{BLATTER}. This indicates that the topological
susceptibility is not a physical quantity in the O(3) nonlinear $\sigma$
model. This observation is supported by results of semiclassical
approximations \cite{SCHWAB,JEVICKI}, where the instanton size distribution is
divergent for small instanton sizes.

In ${\rm CP}^{N-1}$ models with $N>2$, however, semiclassical
approximation indicates that there is no dominance of small
instantons. Furthermore, for $N>3$ dislocations are suppressed even for the
standard lattice action \cite{LUSCHER}.  It should, therefore, be possible, to
determine unambiguously the topological susceptibility in these models. On the
other hand, there were several recent determinations of the topological
susceptibility in the ${\rm CP}^{3}$ model, using different discretizations
of the action and the topological charge
\cite{JANSEN,CAMPOSTRINI,WOLFF,HASENBUSCH,IRVING}; but the results of these
determinations are partly in plain contradiction with each other. The
situation is therefore by no means clear.  We suggest to use the same concepts
and methods in order to determine the topological susceptibility as was
proposed in Ref.~\cite{BLATTER}.

This paper is organized as follows. In Sec. \ref{actionsection} we
construct a classical perfect lattice action for the ${\rm CP}^{3}$ model
along the same lines as in Ref.~\cite{HASENFRATZ}. We define a RG
transformation and determine the FP with analytical and numerical methods. We
give a parametrization of the FP action, that works reasonably well even for
coarse grained fields. In Sec. \ref{chargesection} we define a FP
topological charge along the lines of in Ref.~\cite{BLATTER}. In order to use
the FP charge in numerical simulations, we construct a parametrization of the
dependence of the fine field on a coarse input field. In Sec.
\ref{susceptsection} we discuss the topological susceptibility and the
influence of a lattice regularization on its measurement. Finally, we present
the results of numerical simulations using the FP action and the FP charge.


\section{${\rm \bf CP}^{\bf N-1}$ Models}

Two-Dimensional ${\rm CP}^{N-1}$ models are an important testing ground for
methods in quantum field theories because of their similarities with
four-dimensional non-Abelian gauge theories. Important common properties are
asymptotic freedom, dynamical mass generation, confinement of
non-gauge-invariant states and a nontrivial topology. Despite these common
features, two-dimensional spin models are much easier to handle both
analytically and numerically.

The ${\rm CP}^{N-1}$ model (in the continuum) consists of a $N$-component
complex spin field $z^i(x)$ which satisfies the constraint $\bar{z}(x)\cdot
z(x)$=1.  The action has a global ${\rm SU}(N)$ symmetry and a local $U(1)$
(gauge) symmetry. The gauge field, however, is not independent and can be
expressed in the basic field $z(x)$. Only after quantization, the gauge field
becomes dynamical, and gives rise to a confining potential, as is explicitly
seen in the large-$N$ limit \cite{DADDA}.  In the following it will sometimes
prove useful, not to use the basic field $z(x)$, but the gauge invariant
composite operator $P(x) =z(x) \otimes\bar{z}(x)$.  For $N=2$ the ${\rm
CP}^{N-1}$ model is equivalent to the O(3) nonlinear $\sigma$ model and $P(x)$
can be written in terms of the usual O(3) covariant spins ${\bf S}(x)$ as
$P(x)=\frac{1}{2} ({\sf{1}\kern-.18em\sf{I}\kern+.18em\kern-.18em} + {\bf
\sigma}\cdot{\bf S}(x))$. Here $\bf \sigma$ are the usual Pauli matrices. In
the notation with the composite operators the continuum action for ${\rm
CP}^{N-1}$ assumes the form
\begin{equation}
  N\beta{\cal A}_{\rm cont} = \frac{N\beta}{2} \int d^2x\, \mbox{tr}\left \{
  \partial_{\mu} P(x) \partial_{\mu} P(x) \right \}.
\end{equation}

One has several possibilities to discretize the continuum theory, and to put
it on a lattice \cite{DIVECCHIA}. If one approaches the continuum limit, each
of it will give the same results. We choose for this study a discretization,
which does not use an explicit gauge field. The standard lattice action
without gauge fields assumes the form
\begin{equation}
  N\beta{\cal A}_{\mbox{\tiny ST}} = N\beta \sum_{n,\mu} \left \{ 1-
  |\bar{z}_{n+\hat{\mu}}z_n|^2 \right \}.
\end{equation} 
For $N=2$ this goes over to the standard lattice action of the O(3) nonlinear
$\sigma$ model, which is not the case if one would use a formulation with an
explicit gauge field. The parametrization of the FP action will be a
generalization of the standard action, taking into account couplings between
two spins, that are more distant than nearest neighbor and also multispin
couplings between up to four spins.


\section{Fixed Point Action}
\label{actionsection}

\subsection{Equation for the fixed point action}
We consider the ${\rm CP}^{N-1}$ model on a two-dimensional square
lattice with variables $z_n$ at each lattice site $n$. Then we perform an
exact RG transformation.  For this we split the lattice into $2 \times 2$
blocks $n_B$ and with each block we associate a block spin $\zeta_{n_B}$.  We
define the (gauge invariant) RG transformation as an averaging over the fine
spins in one block, that has the form
\begin{equation}
  e^{-N\beta' {\cal\scriptstyle A}'(\zeta)} = \int_z {\cal N}(z) \,
  \exp\left\{-N\beta{\cal A}(z) + 
  \kappa N\beta \sum_{n_B} \sum_{n \in n_B} |\zeta_{n_B}
  \bar{z}_{n}|^2\right\}.
  \label{RG-trafo} 
\end{equation}
Here, $\kappa$ is a free parameter of the RG transformation, and the
normalizing factor ${\cal N}(z)$ assures, that the partition function does not
change under the RG transformation
\begin{equation}
  {\cal N}(z)^{-1} = \int_{\zeta} \exp\left\{\kappa N\beta \sum_{n_B}
  \sum_{n \in n_B} |\zeta_{n_B} \bar{z}_{n}|^2\right\}.
\end{equation}
The form of the RG transformation in Eq.~(\ref{RG-trafo}) was chosen, such
that for the ${\rm CP}^{1}$ model it corresponds to the one used for the O(3)
nonlinear $\sigma$ model in Ref.~\cite{HASENFRATZ}. In order to compute the
normalizing factor ${\cal N}(z)$, consider the unitary matrix
\begin{equation} 
  M_{n_B} = \sum_{n \in n_B} z_n \otimes \bar{z}_n.
  \label{matrix}
\end{equation}
It is just the sum of the composite operators $P_n$ at every site of a
block. Let the Hermitian matrix $M_{n_B}$ have eigenvalues $\lambda_{n_B}^{i}$
and eigenvectors $w_{n_B}^{i}$.  Because $M_{n_B}$ is gauge invariant, also
its eigenvalues and eigenvectors do not depend on the gauge.  Using these
eigenvalues and eigenvectors, we can rewrite one term in Eq.~(\ref{RG-trafo})
\begin{equation}
  \sum_{n \in n_B} |\zeta_{n_B} \bar{z}_{n}|^2 =
  \bar{\zeta}_{n_B}M_{n_B}\zeta_{n_B} = \sum_{i}
  \lambda_{n_B}^{i} |\zeta_{n_B} \bar{w}^i_{n_B}|^2. 
\end{equation} 
In the limit $\beta \rightarrow \infty$ we perform a saddle point
approximation. We get to lowest order ${\cal N}(z)^{-1} = \exp (\kappa
N\beta\sum_{n_B} \hat{\lambda}_{n_B})$, where $\hat{\lambda}_{n_B}$ is the
largest eigenvalue of $M_{n_B}$.  The FP of the RG transformation
(\ref{RG-trafo}) is determined in this limit by the implicit equation
(FP~equation)
\begin{equation}
  {\cal A}_{\mbox{\tiny FP}}(\zeta) = \min_{\{z\}}\left\{{\cal A}_{\mbox{\tiny
  FP}}(z) + {\cal T}(\zeta,z) \right\}, 
  \label{FP-equation}
\end{equation}
with the transformation kernel 
\begin{equation}
  {\cal T}(\zeta,z) = \kappa \sum_{n_{B}} \left (  \hat{\lambda}_{n_B} - 
\sum_{n \in n_B} |\zeta_{n_B} \bar{z}_{n}|^2 \right ).  
  \label{kernel}
\end{equation}
The kernel ${\cal T}(\zeta,z)$ is gauge invariant in a strong sense: It does
not change under independent gauge transformations on either the coarse {\it
or} on the fine spins. This is more than would be needed: in a general case it
would be sufficient, that the kernel is gauge invariant under a combined gauge
transformation on the coarse {\it and} the fine spins. The gauge invariance of
the kernel assures, that if the action ${\cal A}_{\mbox{\tiny FP}}(z)$ of the
fine field is gauge invariant, then also the action ${\cal A}_{\mbox{\tiny
FP}}(\zeta)$ of the coarse field after the RG transformation is gauge
invariant.

The FP~equation~(\ref{FP-equation}) determines the value of the FP action for
a given input configuration $\{\zeta\}$. One may solve this equation
iteratively, leading to a minimization on a multigrid of lattice configurations
with the configuration $\{\zeta\}$ on the coarsest level, and with $k$ finer
configurations $\{z^{(k)}\}$ on successive levels:
\begin{equation}
  {\cal A}^{(k)}(\zeta) = \min_{\{ z^{(1)}, z^{(2)},\ldots, z^{(k)} \}} 
  \left \{ {\cal A}^{(0)}(z^{(k)}) +
  {\cal T}(z^{(k-1)},z^{(k)}) + \ldots + {\cal T}(\zeta,z^{(1)})
  \right \}.
  \label{iFP-equation}
\end{equation} 
On each successive level the spin configurations become smoother and smoother,
hence one may choose for the action ${\cal A}^{(0)}(z^{(k)})$ on the finest
configuration $\{z^{(k)}\}$ any lattice discretization of the continuum
action. The FP action ${\cal A}_{\mbox{\tiny FP}}(\zeta)$ is then obtained in
the limit $k \rightarrow \infty$ of ${\cal A}^{(k)}(\zeta)$. For practical
purposes, however, only a few levels are needed, and, starting from the
standard action on the lowest level, the FP value is reached soon.  This
iterative method can be used to solve the FP~equation numerically.  One can,
however, make some important statements even without explicitly solving
Eq.~(\ref{FP-equation}). This will be done in the next section.


\subsection{Classical Solutions}
\label{instanton}

As in the O(3) nonlinear $\sigma$ model we can show the following
important statement concerning classical solutions (e.g. instanton solutions).

{\bf Statement}: If the coarse configuration $\{\zeta\}$ satisfies the FP
classical equations of motion (i.e. the classical equations corresponding to
${\cal A}_{\mbox{\tiny FP}}$), and therefore is a local minimum of ${\cal
A}_{\mbox{\tiny FP}}(\zeta)$, then the configuration $\{z(\zeta)\}$ on the
fine lattice, which minimizes the right hand-side of Eq.~(\ref{FP-equation}),
satisfies the equations of motion as well.  In addition, the value of the
action remains unchanged: ${\cal A}_{\mbox{\tiny FP}}(z(\zeta))={\cal
A}_{\mbox{\tiny FP}}(\zeta)$.

{\bf Proof}: 
Since $\{\zeta\}$ is a solution of the FP classical equations of motion, it is
a stationary point of ${\cal A}_{\mbox{\tiny FP}}(\zeta)$, and it satisfies
$\delta{\cal A}_{\mbox{\tiny FP}}/ \delta\zeta = 0$. The transformation kernel
is positive:
\begin{equation}
  {\cal T}_{n_B} = \hat{\lambda}_{n_B}  
  - \bar{\zeta}_{n_B}M_{n_B}\zeta_{n_B} \geq 0. 
\end{equation}
The minimizing configuration $\{z\}$ must fulfill ${\cal T}_{n_B} = 0$,
otherwise we could find another coarse configuration $\{\zeta\}$, that fulfills
this equation and lowers therefore the value of ${\cal A}_{\mbox{\tiny
FP}}(\zeta)$; which would be in contradiction to the assumption, that
$\{\zeta\}$ is a stationary point. Hence we have a (albeit not unique) relation
between the given coarse field $\{\zeta\}$ and the minimizing fine field
$\{z(\zeta)\}$
\begin{equation}
  M_{n_B}\zeta_{n_B} = \hat{\lambda}_{n_B} \zeta_{n_B},
  \label{blocking}
\end{equation}
in other words, $\zeta_{n_B}$ is the eigenvector corresponding to the largest
eigenvalue of the matrix $M_{n_B}$ defined in Eq.~(\ref{matrix}).  ${\cal
T}(\zeta,z)$ is zero (that means, at its absolute minimum) for the minimizing
fine field $\{z(\zeta)\}$, hence $\{z(\zeta)\}$ is also a stationary point of
${\cal A}_{\mbox{\tiny FP}}$ and $\{z(\zeta)\}$ satisfies the classical
FP~equations of motion.  Furthermore, the value of the action is the same:
${\cal A}_{\mbox{\tiny FP}}(z(\zeta))={\cal A}_{\mbox{\tiny FP}}(\zeta)$ q.e.d.

Note, that the reverse statement is not always true. If the fine configuration
$\{z\}$ is a solution of the equations of motion, then the coarse
configuration obtained from Eq.~(\ref{blocking}) is a local minimum, but it
need not be the absolute minimum that has to be found at the right-hand side
of the FP equation~(\ref{FP-equation}). This mechanism prevents the existence
of arbitrarily small instantons on the lattice (see below).

Using the above statement, we can now construct instanton configurations on the
lattice (see also Ref.~\cite{BLATTER} for the corresponding construction in
the O(3) nonlinear $\sigma$ model). We consider instanton configurations of
the continuum on a torus. On a torus, however, there exists no exact
one-instanton configuration \cite{RICHARD}. In order to clearly separate cutoff
effects from finite size effects, we have to fall back upon exact two-instanton
configurations. The two-instanton configurations of ${\rm CP}^{1}$, embedded
into ${\rm CP}^{N-1}$, have the form
\begin{equation}
  z(w)={{\cal N}} \left[\vec{u} + 
  \prod_{i=1}^{2} \frac{\sigma (w-a_i)}{\sigma (w-b_i)} \vec{v} \right],
  \label{inst-cont}
\end{equation}
\[
w =x + iy, \quad \vec{u}=(1,0,0,\ldots), \quad \vec{v}=(0,1,0,\ldots),
\quad \sum_{i=1}^{k} a_i = \sum_{i=1}^{k} b_i,
\]
where $\sigma(n)$ is the Weierstrass $\sigma$ function and the factor ${\cal
N}$ ensures the correct normalization.  The vectors $\vec{u}$ and $\vec{v}$
are $N$-component ${\rm CP}^{N-1}$ vectors which specify the
orientation of the instantons in ``color space''. The four complex parameters
$a_1$, $a_2$, $b_1$, and $b_2$ specify the size of the instantons and their
position and orientation on the torus.  We choose them in the form
\begin{eqnarray}
  a_1 =  \left(L/2 - \rho\right)i, \qquad a_2 = L/2 +
      \left(L/2 + \rho\right)i, \\
  b_1 =  \left(L/2 + \rho\right)i, \qquad b_2 = L/2 +
      \left(L/2 - \rho\right)i, \nonumber
\end{eqnarray}
where $\rho$ is the instanton size and $L$ the size of the torus.  The action
${\cal A}_{\rm cont}$ of the configuration~(\ref{inst-cont}) is equal $2\times
2\pi$ irrespective of the size $\rho$ of the instantons.  Next we want to
construct lattice two-instanton configurations of different sizes. We first
discretize the continuum two-instanton configuration on a very fine lattice, so
that the cutoff effects can safely be neglected.  Then we perform $k$ block
transformations using Eq.~(\ref{blocking}).  Under a block transformation the
size of the instantons is halved.  Choosing the number $k$ of block
transformations and the initial size $\rho$, one can get any final size $\rho\;
2^{-k}$ on a coarse lattice. The above statement shows, that the action
remains the same, unless the size of the instantons is too small, and they
fall through the lattice.

We have numerically performed the above program for the ${\rm CP}^{3}$ model,
and have measured several quantities on the finally blocked configurations.
On the coarse configuration itself we measured the geometric charge, the
standard action, the Symanzik improved action and the parametrization of the
FP action presented in Sec.~\ref{paraction}. Performing a minimization on a
multigrid with three finer levels, we measured the exact FP action and on the
finest level the FP charge (cf. Sec.~\ref{chargesection}).  Because
instantons in ${\rm CP}^{3}$ are embedded ${\rm CP}^{1}$ instantons, it is
clear, that the results presented in Fig.~\ref{inst} are practically identical
to the ones obtained in the O(3) nonlinear $\sigma$ model~\cite{BLATTER}.  The
results show, that it is possible to obtain a parametrization of the FP action
that performs very well for instanton configurations down to the smallest
possible size on the lattice. In contrast, the standard action and also the
Symanzik improved action perform quite bad, especially for small
instantons. It is worth mentioning, that the range of instanton sizes, where
the geometric charge differs from the FP charge, is quite narrow. For
instanton configurations, the geometric charge seems to be as good as the FP
charge. For general configurations created in a Monte Carlo (MC) simulation,
however, there is a noticeable difference between the two charges
(cf. Sec.~\ref{MC}).


\subsection{Parametrization of the FP action}
\label{paraction}

If we want to use the FP action in practical numerical simulations, we must
construct a parametrization, that consists of a not too large set of local
operators. Our parametrization has the form
\begin{equation}
  {\cal A}_{\mbox{\tiny FP}}^{\rm par}(z) = -{1\over 2} \sum_{n,r} \rho(r)
  \theta_{n,n+r}^2 + \sum_{n_i,n_j,\ldots} \mbox{ coupling }\times \mbox{
  products of } \theta_{n_i,n_j}^2, \label{parametrization}
\end{equation}
where $\theta_{n_i,n_j} = \arccos\left(\left| \bar{z}_{n_i}
z_{n_j}\right|\right)$ is the angle between two spins. Note, that in ${\rm
CP}^{N-1}$ models the maximal angle between two spins is $\pi /2$; $z$ and
$-z$ is the same spin, different only by a gauge transformation. There are two
reasons why it is useful to use the angle $\theta_{n_i,n_j}$ instead of
$\left(1-|\bar{z}_{n_i} z_{n_j}|^2\right)$: For solutions of the equations of
motion of the form $z(n)=(\cos\theta n_0,\,\sin\theta n_0,\, 0,\ldots )$, the
$\theta$ dependence of the action is exactly $\theta^2$. Moreover, if one
rotates a single spin in a trivial background with all spins pointing in the
same direction, the $\theta$ dependence of the action is more like $\theta^2$
rather than $1-\cos^2(\theta)$. Figure~\ref{onespin} shows how well the
minimized FP~action is approximated even by the two first terms (nearest
neighbor, diagonal) of the lowest order of the parametrization in
Eq.~(\ref{parametrization}). In contrast, the standard action performs very
bad for large angles.

We can calculate the coefficients $\rho(r)$ of the lowest order analytically. 
The coefficients of the higher orders can be determined in a numerical fitting
procedure.

\subsubsection{Lowest order, determination of $\rho$}

The analytic result for the lowest order term in the FP action will be valid
for all $N$, as long as $N>1$.  Consider on the coarse level a smooth
configuration, which is weakly fluctuating around the first direction
\begin{equation}
  \zeta_{n_B}=\left( {\sqrt{1-|X_{n_B}|^2}
  \atop X_{n_B}} \right),
  \label{coarse_expansion}
\end{equation}
where $X_{n_B}$ has $(N-1)$ components, and $|X_{n_B}|\ll 1$. With this choice
we have fixed the gauge by imposing, that the first component is real and
positive. Then the minimizing fine configuration will also fluctuate around
the first direction and, using the same gauge fixing prescription, we make the
ansatz
\begin{equation}
  z_{n}=\left( {\sqrt{1-|V_{n}|^2}
  \atop V_{n}} \right),
  \label{fine_expansion}
\end{equation}
with $|V_{n}|\ll 1$. Putting these expansions and the parametrization of the
action~(\ref{parametrization}) in the FP~equation~(\ref{FP-equation}) and
keeping only terms up to quadratic order in $X$ and $V$, we get the equation
\begin{eqnarray}
\lefteqn{{1\over2}\sum_{n_B,r_B} \rho(r_B)
{\rm Re}(\bar{X}_{n_B}X_{n_B+r_B})=} \nonumber    \\ 
 & & \min_{ \{ V\} } 
\left\{ 
{1\over2}\sum_{n,r} \rho(r){\rm Re}(\bar{V}_nV_{n+r})
+2\kappa\sum_{n_B}
\left| X_{n_B}- {1\over 4}\sum_{n\in n_B} V_n \right|^2
\right\} \,.
\label{quadratic-approx}
\end{eqnarray}
This equation can be solved for $\rho$ most easily by using the same technique
as described in Ref.~\cite{HASENFRATZ}. Taking into account the fact, that for
the ${\rm CP}^{1}$ model the RG transformation goes over to the one used for
the O(3) nonlinear $\sigma$ model, it is not astonishing, that the resulting
$\rho$ is exactly the same as in Ref.~\cite{HASENFRATZ}. Consequently, in
order to have a most local action, we will also choose the free parameter
$\kappa=2$ in the RG transformation Eq.~(\ref{RG-trafo}).

\subsubsection{Higher orders for the ${\rm CP}^{3}$ model}

In order to have a reasonably good parametrization of the lattice action which
performs well also for coarse configurations, we include higher order terms in
the parametrization~(\ref{parametrization}). The coefficients of these higher
order terms will depend on the chosen model, and are different for each $N$.
We determined them numerically for the ${\rm CP}^{3}$ model. To do this, we
first produced about 300 configurations of lattice size 3 with a Monte Carlo
program using the standard action. The configurations ranged from ones with
small actions and fluctuations with small amplitudes to strongly fluctuating
ones.  We also included some two-instanton configurations with radii of the
order of one lattice spacing (cf. Sec.~\ref{instanton}). For every
configuration we calculated the value of the FP~action by minimizing the
FP~equation on a multigrid. Then we determined the coefficients of 30 higher
order operators chosen because of their locality. Their value was determined
so that the average difference between minimized and parametrized FP~action
was minimal.

Figure~\ref{scatter} illustrates the quality of the parametrization. The
relative deviation from the minimized FP~action is only large for very coarse
configurations with large actions. For comparison, we also plot the standard
action for the same configurations. Note, that here the relative deviation of
the standard action from the FP action is not small, even for configurations
with small amplitude fluctuations.

The resulting 32 coefficients of the parametrization~(\ref{parametrization})
are given with a graphical notation of the corresponding operators in
Table~\ref{couplings}. A line with two dots 
\begin{picture}(22,6)(0,0) 
\put(2,3){\circle*{2.5}} 
\put(20,3){\circle*{2.5}}     
\put(2,3){\line(1,0){18}}     
\end{picture} 
means that the angle $\theta_{n_1,n_2}^2$ between the two spins at positions
$n_1$ and $n_2$ of the dots enters into the parametrization. A graph
consisting of several lines represents just the multiplication of the
corresponding angles.  Coefficients No. 1 and No. 4 are the only analytically
calculated ones, all the other couplings were determined by the fitting
procedure.  The locality of the action is expressed in the fact, that it is
possible to construct a good parametrization with operators consisting of
spins, that lie only within a $2\times 2$ section of the whole lattice. On the
other hand, the parametrization seems to be not as local as the one found in
the O(3) nonlinear $\sigma$ model. Some couplings of operators of higher order
have quite a large value. Nevertheless, one should not give too much attention
to the actual value of the couplings. A large coupling for a higher order
operator does not mean, that the action is less local. Some combinations of
couplings can be changed without much affecting the quality of the
parametrization. In principle one would not need to keep all 32 operators in
the parametrization, we just did so for reasons of completeness. Using 32
operators for an action in a MC simulation may seem to be a lot. However,
compared with the expected benefit of using such an action, the additional
computational effort needed seems to be reasonable.


\section{Fixed Point Topological Charge}
\label{chargesection}

A feature of ${\rm CP}^{N-1}$ models common with gauge field models is the
existence of topologically nontrivial solutions. In the continuum theory the
topological charge $Q$ may be defined as the integral
\begin{equation}
  Q = \frac{i}{2\pi} \int d^2x\, \varepsilon_{\mu\nu} \, tr \left[
  P(x)\partial_{\mu}P(x)\partial_{\nu}P(x) \right].
\end{equation}
The action is related to the charge through the inequality
\begin{equation}
  {\cal A}_{\rm cont} \geq 2\pi\, |Q|.
  \label{inequality}
\end{equation}
For instantons the equality holds. They minimize the action, and are therefore
solutions of the equations of motion.

On the lattice, however, this concept breaks down because continuity is lost.
In the continuum the topological sectors are clearly separated, but on the
lattice one may continuously transform a field from one topological sector to
another.  Furthermore, when using the standard lattice action, the scale
invariance of instanton solutions is violated. The action decreases with
decreasing instanton size, and configurations that violate the
inequality~(\ref{inequality}) -- so called ``dislocations'' -- are possible.

In Refs.~\cite{HASENFRATZ,BLATTER} it was shown that the FP action admits
stable instanton solutions. Furthermore, it was shown how to construct a
correct charge -- the FP charge -- which does not admit dislocations. In this
section we will proceed along the same lines as in these works, in order
to construct the FP charge for the ${\rm CP}^{3}$ model.


\subsection{Definition of the FP charge}

We define the FP charge by means of the iterated solution of the
FP~equation~(\ref{iFP-equation}). Under a RG transformation in the limit
$\beta \rightarrow \infty$, an operator ${ \cal O}(z)$ transforms into ${\cal
O}'(\zeta)$ on the coarse lattice as
\begin{equation}
  {\cal O}'(\zeta) = {\cal O}(z(\zeta)),
\end{equation}
where the spin configuration $\{z(\zeta)\}$ is the solution of the
FP~equation~(\ref{FP-equation}). The FP topological charge is obtained in the
limit of infinitely many RG transformations:
\begin{equation}
  Q_{\mbox{\rm \tiny FP}}(\zeta) = \lim_{k \rightarrow \infty}
  Q(z^{(k)}(\zeta)),
\end{equation}
where $\{ z^{(k)}\}$ is the solution of the iterated
FP~equation~(\ref{iFP-equation}) on the lowest level in a $k$ level multigrid.
As the configurations get smoother on each successive level, one may choose
for the charge $Q$ on the lowest level any lattice discretization of the
topological charge. In this paper we will use the geometric charge
\cite{BERG}, because it is stable against small variations of the field, if the
field is sufficiently smooth, and because it always gives an integer
number.

In Ref.~\cite{BLATTER} it was shown, that the combination FP action and
FP topological charge always obeys the inequality
\begin{equation}
  {\cal A}_{\mbox{\rm \tiny FP}}(\zeta) \geq 2 \pi \, | Q_{\mbox{\rm \tiny
  FP}}(\zeta) |.
\end{equation}
Hence there are no dislocations.

In numerical simulations it is very time consuming to minimize a multigrid for
every configuration. One needs a parametrization of the solution
$\{z(\zeta)\}$ of the FP~equation. This will be done in the next section.


\subsection{Fixed point field}
\label{fpfield}

The FP field is the fine field $z^{(k)}$ in the multigrid solution of the
iterated FP~equation~(\ref{iFP-equation}) as $k$ goes to infinity.  If the
functional dependence of the solution on the first fine level $z^{(1)}$ on
$\zeta$ is known, the FP field can be evaluated by iteration. Below we give an
expression for the field $z^{(1)}=z^{(1)}(\zeta)$.  (In the limit $k
\rightarrow \infty$ the solution $\{ z^{(1)} \}$ of the iterated FP~equation
is identical to the solution $\{ z \}$ of the FP~equation.)

We make the expansions (\ref{coarse_expansion}) and (\ref{fine_expansion}) for
the coarse and the fine configuration. Then the solution of the FP~equation
leads in lowest order to the relation
\begin{equation}
  V_n = \sum_{n_B} \alpha (n,n_B) X_{n_B}
\end{equation}
with $\alpha$ the same as for the O(3) $\sigma$ model~\cite{HASENFRATZ,%
DEGRAND,BLATTER,BELL}. 

To obtain this result, we have fixed the gauge in a specific way. However, we
want a relation between $\{z\}$ and $\{\zeta\}$, that does not depend on the
choice of the gauge fixing prescription. Furthermore, for coarse
configurations we have to include higher order terms, that parametrize the
dependence of $\{z\}$ on $\{\zeta\}$.  For this we construct the gauge
invariant composite operators $P_{n_B} = \zeta_{n_B} \otimes
\bar{\zeta}_{n_B}$ on the coarse lattice. On the fine lattice we build
matrices $Q_n$ by, summing over the coarse composite operators including the
next to leading order terms:
\begin{equation}
  Q_n = \sum_{n_B} \alpha
       (n,n_B)P_{n_B} + \!\! \sum_{n_{\mbox{\tiny B}}
       \atop m_B^{},m_{B}'} \!\! \beta(n,n_{\mbox{\tiny
       B}},m_{B}^{},m_{\mbox{\tiny B}}') 
       \theta^2_{m_B^{},m_B'} P_{n_B},
  \label{parametfield}
\end{equation}
and define the fine field variable $z_n$ as the eigenvector of $Q_n$ with
largest eigenvalue. In next order enters the angle $\theta^2_{m_B^{},m_B'}$
between the coarse spins at sites $m_B$ and $m_B'$, respectively. In order to
determine the coefficients $\beta$, we used the same $\approx 300$
configurations like for the parametrization of the action, minimized the
FP~equation~(\ref{FP-equation}) and stored the resulting fine lattices. The
coefficients were then determined by minimizing the difference between the
minimized fine spins and the parametrization (\ref{parametfield}).

The numerical values of the coefficients $\alpha$ and $\beta$ are given in
Table~\ref{alphabeta} with a graphical notation of the corresponding
operators. We chose a set of 17 operators mainly because of their
compactness. Numbers 1--6 are the analytically determined coefficients
$\alpha$, Nos. 7--17 are the numerically determined coefficients $\beta$.  The
meaning of the graphical notation of the operators is the following: The
dashed lines represent a $3\times 3$ section of the coarse lattice grid. The
cross
\begin{picture}(8,6)(0,0)  
\put(4,3){\line(1,0){3.5}}
\put(4,3){\line(0,1){3.5}}        
\put(4,3){\line(-1,0){3.5}}       
\put(4,3){\line(0,-1){3.5}}
\end{picture} 
in between indicates the position $n$ of the matrix $Q_n$ in
Eq.~(\ref{parametfield}). The little square
\begin{picture}(8,6)(0,0)         
\put(2,1){\framebox(4,4)}         
\end{picture} 
denotes the position $n_{\mbox{\tiny B}}$ of the coarse composite operator
$P_{n_{\mbox{\tiny B}}}\,$, the two connected dots
\begin{picture}(22,6)(0,0) 
\put(2,3){\circle*{2.5}}          
\put(20,3){\circle*{2.5}}         
\put(2,3){\line(1,0){18}}
\end{picture} 
are the positions $m_{\mbox{\tiny B}}$ and $m_{\mbox{\tiny B}}'$ of the spins
whose angle $\theta^2_{ m_{\mbox{\tiny B}}^{}, m_{\mbox{\tiny B}}'}$ enters
into the parametrization.  Graphs obtained by trivial symmetry transformations
are not drawn separately.


\section{Topological Susceptibility}
\label{susceptsection}
The topological susceptibility is defined as the ratio
\begin{equation}
  \chi_t = \frac{\langle Q^2\rangle}{V},
\end{equation}
where $Q$ is the topological charge and $V$ is the space--time volume. In
${\rm CP}^{N-1}$ models it is a dimension two quantity, that vanishes to
all orders in the weak coupling expansion. From the perturbative
renormalization group it is expected to scale according to the two loop
$\beta$ function
\begin{equation} 
  \chi_t \propto \left (\frac{1}{2} \pi N\beta \right )^{4/N} 
  \exp(-  \pi N\beta) \qquad ( \beta \rightarrow \infty).
  \label{eq-scaling}
\end{equation}
In order to check the continuum limit, a scaling behavior is more important to
observe than the above asymptotic scaling.  One additionally measures a second
quantity, e.g., the correlation length $\xi$, and builds the dimensionless
product $\chi_t\, \xi^2$, which should go to a constant in the continuum limit
$\xi \rightarrow \infty$.

In general, cutoff effects can originate from two sources: from the
discretization of the action and from the discretization of an operator. The
topological charge is an operator, that strongly exhibits lattice cutoff
effects, as was demonstrated in Sec.~\ref{instanton}: instantons with a
radius smaller than about $0.7a$ fall through the lattice and are lost in a
Monte Carlo simulation. This fraction is large at correlation lengths of the
same order as the lattice spacing, and gets smaller as the correlation length
grows.  One therefore expects a scaling violation, which is determined by the
small instanton size distribution.

One may estimate the behavior of the expected scaling violation by using the
results of a semiclassical expansion, and performing a kind of dilute
instanton gas approximation. In ${\rm CP}^{3}$ the probability density to
find a field configuration with topological charge $Q=1$ in a sphere with
radius $R$ is \cite{SCHWAB}
\begin{equation}
  D_1(R) = K_1(4) \,\Lambda_{\mbox{\tiny MS}}^4 R^2 
  \ln^2(\Lambda_{\mbox{\tiny MS}} R) \qquad (R \rightarrow 0)
\end{equation}
where $\Lambda_{\mbox{\tiny MS}}$ is the perturbative $\Lambda$ parameter and
$K_1(4)=36.2995$ results from the integration over the instanton
parameters. This result is valid for small volumes, so we can use it to
estimate the probability of loosing a charge-one configuration that falls
through a small lattice mesh.

We denote the charge measured on the lattice with $Q$ and the lost charge with
$q$. If we assume that $Q$ and $q$ are independent, then the topological
susceptibility measured on the lattice is
\begin{equation}
  \chi_t^{\rm lat} = \frac{\langle Q^2\rangle}{V} = \chi_t^{\rm cont} -
  \frac{\langle q^2\rangle}{V}.
\end{equation}
We make a kind of dilute gas approximation by assuming that within each
lattice mesh there can be independently an instanton or an antiinstanton.  The
result is
\begin{equation}
  \frac{\langle q^2\rangle}{V} = 2D_1.
\end{equation}
Identifying $R$ with the minimal size $\rho=ca$ of an instanton on the
lattice ($c\simeq 0.7$) provides
\begin{equation}
  \chi_t^{\rm lat} = \chi_t^{\rm cont} - 2K_1(4) \,\Lambda_{\mbox{\tiny MS}}^4
  c^2 a^2 \ln^2(\Lambda_{\mbox{\tiny MS}} ca), \qquad (a \rightarrow 0)
\label{dilute}
\end{equation}
which can be compared with the results of a Monte Carlo simulation.


\section{Numerical Results}
\label{MC}

We performed Monte Carlo simulations in the ${\rm CP}^{3}$ model using the
parametrized FP action given in Eq.~(\ref{parametrization}). In order to
reduce critical slowing down, we implemented a hybrid overrelaxation algorithm
similar to the one described in Ref.~\cite{WOLFF}. The mass gap $m=\xi^{-1}$
was obtained from the long distance fall off of the correlation function
projected to zero spatial momentum. We made a minimal $\chi^2$ fit to the
correlation function with the function $c(e^{-mx}+e^{-m(L-x)})$ in the
interval $x \in [x_{\rm min},L/2]$ for different $x_{\rm min}$. The value of
$m$ is taken at an expected plateau at $x_{\rm min} \gtrsim \xi$.

In order to avoid finite size effects, one has to make measurements in large
enough volumes. We made measurements for all $\beta$ values in volumes with
the ratio $L/\xi \simeq 5.5 - 6$ kept approximately constant. These volumes
are usually large enough to totally avoid any finite size effects. In ${\rm
CP}^{N-1}$ models, however, the basic $z$ particles are subject to a
confining potential. Thus the resulting bound states may have a radius that is
larger than the correlation length. For this reason we also made measurements
for some $\beta$ values in even larger volumes (with ratios $L/\xi$ up to 12).

The masses that are determined in the ``small'' volume $L/\xi \simeq 5.6$ are
about $3 \%$ below the ones measured in ``infinite'' volume, as can be seen in
Table~\ref{mcres}. The magnitude of this finite size effect fits nicely with
the one observed in Refs.~\cite{CAMPOSTRINI,ROSSI}. Furthermore, we
observed, that in the large volumes the determination of the mass gap showed a
nice plateau behavior. Such a plateau was sometimes not very clearly seen in
the small volumes. Nevertheless, we may use the results obtained in the small
volume in order to look for a scaling behavior in $\chi_t \xi^2$. The
actual values are spoiled by finite-size effects, but since $L/\xi$ is
constant, these effects are the same for every point.

We checked asymptotic scaling of the mass gap according to the perturbative
lattice scale in two-loop approximation
\begin{equation}
  \Lambda^{(2)}_L = \left (\frac{1}{2} \pi N\beta \right )^{2/N} 
  \exp\left (- \frac{1}{2} \pi N\beta \right ).
\end{equation}
In Fig.~\ref{asymptotic} we show the ratio $m/\Lambda^{(2)}_L$ versus the
correlation length, with masses determined in ``small'' volumes $L/\xi \simeq
5.6$ and also some measurements in ``large'' volumes. The most striking
observation is that the ratio approaches a constant value $m/\Lambda^{(2)}_L =
8.1(1)$ (this value is obtained in ``large'' volumes), and that its final
value is attained already at a quite short correlation length. Such a
precocious asymptotic scaling has not been stated for other lattice
actions. With the standard action, for example, one does not even see an
asymptotic scaling at correlation lengths $\xi \sim
50$~\cite{WOLFF,HASENBUSCH}. Note that the asymptotic scaling behavior for the
FP action is a purely phenomenological observation that has not been expected
theoretically. The FP action was constructed nonperturbatively in order to
reduce cutoff effects.  That the mass gap scales with the perturbative
$\Lambda$ parameter is just an additional, unexpected benefit.  We furthermore
observe, that the value of the ratio $m/\Lambda^{(2)}_L$ is remarkably
small. From this we conclude, that $\Lambda_{\rm FP}$ is much closer to the
continuum scale than $\Lambda_{\rm ST}$.

We determined the topological charge, using both the geometric definition and
the definition of the FP charge given in Sec.~\ref{chargesection}. For the
measurement of the FP charge we used the geometric charge on a finer lattice
of the multigrid with the Monte Carlo generated lattice as coarsest level.  In
order to determine the configuration $z(\zeta)$ on the first finer level, one
can either minimize the FP~equation (which is very time consuming) or use the
parametrization of the dependence on $\zeta$, given in
Sec.~\ref{fpfield}. We denote the corresponding charges $Q_{\rm coarse}$
for the geometric charge and, e.g., $Q_{\rm par\,1.\,level}$ for the charge
measured on the first finer level using the parametrization of the fine field.

For some $\beta$ values we compared the results of using the parametrization
on a finer level and of minimizing on a multigrid. The results were found to
be consistent within the statistical errors, as is reported in
Table~\ref{mcres}.  This shows that the parametrization performs well for
typical configurations occurring in a Monte Carlo simulation.  We wanted to be
sure, that it is sufficient to measure the charge only on the first finer
level. As a test we calculated for one $\beta$ value also the charge on the
second finer level.  The result presented in Table~\ref{mcres} shows, that the
values on the first and the second finer level were found to be consistent
within the statistical errors.  This means that the process of going to a
lower level is already stable at that stage, and that it is sufficient to
calculate the fine field only on the first finer level. This is not
unexpected, since we use a geometric definition of the topological charge at a
lower level. Therefore, we always get an integer number for the charge, and
the charge has to change abruptly when going to a finer level. The field on
the first finer level is then (in most cases) smooth enough to yield already
the FP topological charge.

Table~\ref{mcres} shows the effect of using different definitions of the
lattice topological charge. On the coarse level the value of $\langle Q^2_{\rm
coarse} \rangle$ is higher than the one obtained with the FP charge (on the
first finer level). This can be explained with dislocations which contribute to
the geometric charge. They have the effect that one overestimates the amount
of topological excitation.

In Fig.~\ref{suscresults} we show the results of the scaling test for the
dimensionless quantity $\chi_t \xi^2$, measured in volumes with a ratio
$L / \xi \simeq 5.6$. One clearly sees the expected raise at small correlation
lengths, which is due to lost small instantons in this region. At correlation
length $\xi \simeq 10$ this effect is already saturated.  The measurements in
``small'' volumes using the FP charge (i.e., $Q_{\rm par\,1.\,level}$) show the
expected scaling plateau at a value of about $\chi_t \xi^2 \simeq 0.071$
(c.f. Fig.~\ref{suscresults}). The measurements using the geometric definition
of the topological charge give somewhat larger values but they seem to
converge to the value obtained from the FP topological charge for larger
correlation lengths -- these measurements do not show the same nice scaling
behavior as the measurements using the FP charge. This is also what is to be
expected in the ${\rm CP}^{3}$ model if dislocations contribute to the
geometric charge. In the continuum limit the effect of these dislocations is
suppressed, but they still contribute at the considered correlation lengths.

We can now use the four measurements in ``large'' volumes, with $Q_{\rm
par\,1.\,level}$ for the topological charge and perform a fit to the data
with Eq.~(\ref{dilute}), in order to extrapolate to the continuum value. We 
obtain the value $\chi_t \xi^2 = 0.070(2)$.

Let us compare our numerical value with previous determinations and
with results from the large $N$ expansion~\cite{LUSCHER}. To leading order
in $1/N$ one gets
\begin{equation}
\chi_t \xi^2 = \frac{3}{4\pi N} + O(1/N^{5/3}) \simeq 0.06 \qquad
({\rm for}\, N=4). 
\label{largeN}
\end{equation}
This value is comparable with the value we got at the scaling plateau.
However, the correction to the leading order is large in
Eq.~(\ref{largeN}) and the agreement with numerical results occurs at chance at
$N=4$. The numerical results of Ref.~\cite{CAMPOSTRINI} (who use actions
with an explicit gauge field) -- quoting $\chi_t \xi^2 \simeq 0.06$ with
an uncertainty of $10-20\%$ -- are in quite good correspondence with our
result. Measurements using the standard action and the geometric definition of
the topological charge (Refs.~\cite{WOLFF,HASENBUSCH}) lead to a value
roughly twice as large as ours, but with large scaling violations at large
correlation lengths -- this is probably the effect of dislocations which still
contribute at the considered correlation lengths.


\section{Conclusions}
\label{conclusions}

In ${\rm CP}^{N-1}$ models it is possible to define a FP action and a FP
topological charge. The FP action is the fixed point of an exact RG
transformation. It is a classically perfect action and possesses scale
invariant instanton solutions. The definition of the FP topological charge is
based on the FP field operator. Both, the FP action and the FP field can be
evaluated to any precision desired on a sufficiently large multigrid. It can
be shown that the FP topological charge together with the FP action have no
lattice defects. 

It is profitable to use the FP action and the FP topological charge in
numerical simulations. For this purpose, we have parametrized for the ${\rm
CP}^{3}$ model the FP action and the field solution of the FP equation which
is iterated to obtain the FP field. We used these parametrizations in MC
simulations of the ${\rm CP}^{3}$ model. We find two main results. First, the
mass gap unexpectedly scales according to the perturbative lattice scale.
Second, the dimensionless quantity $\chi_t \xi^2$ raises as expected at small
correlation lengths, and reaches a scaling plateau at $\chi_t \xi^2 =
0.070(2)$.

For the future, it would be interesting and profitable to use the parametrized
FP action for further MC simulations, for example in order to investigate the
spectrum of ${\rm CP}^{N-1}$ models.

\acknowledgments

\noindent
I would like to thank P.~Hasenfratz, F.~Niedermayer, M.~Blatter,
U.-J.~Wiese, A.~Papa, and P.~Kunszt for helpful discussions.  This work was
supported in part by the Schweizerischer Nationalfonds.


\begin{figure}
\caption{Actions and charge of instantons with radii of the order of one
lattice spacing.}
\label{inst}
\end{figure}

\begin{figure}
\caption{Actions of configurations with one spin rotated against a trivial
background.}
\label{onespin}
\end{figure}

\begin{figure}
\caption{Scatter plot of the actions of the configurations used for the fit.}
\label{scatter}
\end{figure}

\begin{figure}
\caption{Asymptotic scaling test for the mass in the ${\rm CP}^{3}$ model.}
\label{asymptotic}
\end{figure}

\begin{figure}
\caption{Scaling test for the topological susceptibility. These are the
results measured in ``small'' volumes $L / \xi \simeq 5.6$. The larger error
bars at some results from the first finer level are due to the fact that we
performed less measurements where we also measured the FP charge.}
\label{suscresults}
\end{figure}

\newpage

\begin{table}[h]\centering 
\setlength{\unitlength}{0.7mm}
\begin{tabular}{rcr rcr rcr}
No. & Type & Coupling & No. & Type & Coupling
& No. & Type & Coupling \\
\cline{1-9}
1 &
\begin{picture}(18,10)(0,3)
\put(5,5){\circle*{2}}
\put(15,5){\circle*{2}}
\put(5,5){\line(1,0){10}}
\end{picture} & $-0.61884$ & 

2 &
\begin{picture}(18,10)(0,3)
\put(5,5){\circle*{2}}
\put(15,5){\circle*{2}}
\put(5,4.5){\line(1,0){10}}
\put(5,5.5){\line(1,0){10}}
\end{picture} & $-0.05381$ & 

3 &
\begin{picture}(18,10)(0,3)
\put(5,5){\circle*{2}}
\put(15,5){\circle*{2}}
\put(5,4.1){\line(1,0){10}}
\put(5,5){\line(1,0){10}}
\put(5,5.9){\line(1,0){10}}
\end{picture} & $0.20023$ \\
 & & & & & & & & \\

4 &
\begin{picture}(18,14)(0,5)
\put(5,1){\circle*{2}}
\put(15,11){\circle*{2}}
\put(5,1){\line(1,1){10}}
\end{picture} & $-0.19058$  & 

5 &
\begin{picture}(18,14)(0,5)
\put(5,1){\circle*{2}}
\put(15,11){\circle*{2}}
\put(5,0.5){\line(1,1){10}}
\put(5,1.5){\line(1,1){10}}
\end{picture} & $-0.01892$  & 

6 &
\begin{picture}(18,14)(0,5)
\put(5,1){\circle*{2}}
\put(15,11){\circle*{2}}
\put(5,0.1){\line(1,1){10}}
\put(5,1.9){\line(1,1){10}}
\put(5,1){\line(1,1){10}}
\end{picture} & $-0.06735$ \\
 & & & & & & & & \\

7 &
\begin{picture}(18,14)(0,5)
\put(5,1){\circle*{2}}
\put(15,1){\circle*{2}}
\put(15,11){\circle*{2}}
\put(5,1){\line(1,0){10}}
\put(5,1){\line(1,1){10}}
\end{picture} & $0.01455$  & 

8 &
\begin{picture}(18,14)(0,5)
\put(5,1){\circle*{2}}
\put(15,1){\circle*{2}}
\put(15,11){\circle*{2}}
\put(5,0.5){\line(1,0){10}}
\put(5,1.5){\line(1,0){10}}
\put(5,1){\line(1,1){10}}
\end{picture} & $-0.25328$  & 

9 &
\begin{picture}(18,14)(0,5)
\put(5,1){\circle*{2}}
\put(15,1){\circle*{2}}
\put(15,11){\circle*{2}}
\put(5,1){\line(1,0){10}}
\put(5,0.5){\line(1,1){10}}
\put(5,1.5){\line(1,1){10}}
\end{picture} & $0.07099$ \\
 & & & & & & & & \\

10 &
\begin{picture}(18,14)(0,5)
\put(5,1){\circle*{2}}
\put(15,1){\circle*{2}}
\put(15,11){\circle*{2}}
\put(5,0.5){\line(1,0){10}}
\put(5,1.5){\line(1,0){10}}
\put(5,0.5){\line(1,1){10}}
\put(5,1.5){\line(1,1){10}}
\end{picture} & $0.04334$ &

11 &
\begin{picture}(18,14)(0,5)
\put(5,1){\circle*{2}}
\put(15,1){\circle*{2}}
\put(15,11){\circle*{2}}
\put(5,1){\line(1,0){10}}
\put(15,1){\line(0,1){10}}
\end{picture} & $0.02704$  & 

12 &
\begin{picture}(18,14)(0,5)
\put(5,1){\circle*{2}}
\put(15,1){\circle*{2}}
\put(15,11){\circle*{2}}
\put(5,0.5){\line(1,0){10}}
\put(5,1.5){\line(1,0){10}}
\put(15,1){\line(0,1){10}}
\end{picture} & $-0.12660$ \\ 
& & & & & & & & \\

13 &
\begin{picture}(18,14)(0,5)
\put(5,1){\circle*{2}}
\put(15,1){\circle*{2}}
\put(15,11){\circle*{2}}
\put(5,0.5){\line(1,0){10}}
\put(5,1.5){\line(1,0){10}}
\put(14.5,1){\line(0,1){10}}
\put(15.5,1){\line(0,1){10}}
\end{picture} & $0.06787$ &

14 &
\begin{picture}(18,14)(0,5)
\put(5,1){\circle*{2}}
\put(15,1){\circle*{2}}
\put(15,11){\circle*{2}}
\put(5,1){\line(1,0){10}}
\put(5,1){\line(1,1){10}}
\put(15,1){\line(0,1){10}}
\end{picture} & $0.18327$  & 

15 &
\begin{picture}(18,14)(0,5)
\put(5,1){\circle*{2}}
\put(15,1){\circle*{2}}
\put(15,11){\circle*{2}}
\put(5,0.5){\line(1,0){10}}
\put(5,1.5){\line(1,0){10}}
\put(5,1){\line(1,1){10}}
\put(15,1){\line(0,1){10}}
\end{picture} & $0.13297$ \\
 & & & & & & & & \\

16 &
\begin{picture}(18,14)(0,5)
\put(5,1){\circle*{2}}
\put(15,1){\circle*{2}}
\put(15,11){\circle*{2}}
\put(5,1){\line(1,0){10}}
\put(5,0.5){\line(1,1){10}}
\put(5,1.5){\line(1,1){10}}
\put(15,1){\line(0,1){10}}
\end{picture} & $-0.28036$ &

17 &
\begin{picture}(18,14)(0,5)
\put(5,1){\circle*{2}}
\put(15,1){\circle*{2}}
\put(15,11){\circle*{2}}
\put(5,11){\circle*{2}}
\put(5,1){\line(1,0){10}}
\put(5,11){\line(1,0){10}}
\end{picture} & $-0.00174$  & 

18 &
\begin{picture}(20,14)(0,5)
\put(5,1){\circle*{2}}
\put(15,1){\circle*{2}}
\put(15,11){\circle*{2}}
\put(5,11){\circle*{2}}
\put(5,0.5){\line(1,0){10}}
\put(5,1.5){\line(1,0){10}}
\put(5,11){\line(1,0){10}}
\end{picture} & $0.26017$  \\
& & & & & & & & \\

19 &
\begin{picture}(18,14)(0,5)
\put(5,1){\circle*{2}}
\put(15,1){\circle*{2}}
\put(15,11){\circle*{2}}
\put(5,11){\circle*{2}}
\put(5,0.5){\line(1,0){10}}
\put(5,1.5){\line(1,0){10}}
\put(5,10.5){\line(1,0){10}}
\put(5,11.5){\line(1,0){10}}
\end{picture} & $0.11006$ &

20 &
\begin{picture}(18,14)(0,5)
\put(5,1){\circle*{2}}
\put(15,1){\circle*{2}}
\put(15,11){\circle*{2}}
\put(5,11){\circle*{2}}
\put(5,1){\line(1,1){10}}
\put(5,11){\line(1,-1){10}}
\end{picture} & $0.01396$  & 

21 &
\begin{picture}(18,14)(0,5)
\put(5,1){\circle*{2}}
\put(15,1){\circle*{2}}
\put(15,11){\circle*{2}}
\put(5,11){\circle*{2}}
\put(5,0.5){\line(1,1){10}}
\put(5,1.5){\line(1,1){10}}
\put(5,11){\line(1,-1){10}}
\end{picture} & $0.09222$ \\ 
 & & & & & & & & \\

22 &
\begin{picture}(18,14)(0,5)
\put(5,1){\circle*{2}}
\put(15,1){\circle*{2}}
\put(15,11){\circle*{2}}
\put(5,11){\circle*{2}}
\put(5,0.5){\line(1,1){10}}
\put(5,1.5){\line(1,1){10}}
\put(5,10.5){\line(1,-1){10}}
\put(5,11.5){\line(1,-1){10}}
\end{picture} & $-0.02530$ &
 
23 &
\begin{picture}(18,14)(0,5)
\put(5,1){\circle*{2}}
\put(15,1){\circle*{2}}
\put(15,11){\circle*{2}}
\put(5,11){\circle*{2}}
\put(5,1){\line(1,0){10}}
\put(15,1){\line(0,1){10}}
\put(15,1){\line(-1,1){10}}
\end{picture} & $0.52163$  & 

24 &
\begin{picture}(18,14)(0,5)
\put(5,1){\circle*{2}}
\put(15,1){\circle*{2}}
\put(15,11){\circle*{2}}
\put(5,11){\circle*{2}}
\put(5,0.5){\line(1,0){10}}
\put(5,1.5){\line(1,0){10}}
\put(15,1){\line(0,1){10}}
\put(15,1){\line(-1,1){10}}
\end{picture} & $-0.05146$ \\ 
 & & & & & & & & \\

25 &
\begin{picture}(18,14)(0,5)
\put(5,1){\circle*{2}}
\put(15,1){\circle*{2}}
\put(15,11){\circle*{2}}
\put(5,11){\circle*{2}}
\put(5,1){\line(1,0){10}}
\put(15,1){\line(0,1){10}}
\put(15,0.5){\line(-1,1){10}}
\put(15,1.5){\line(-1,1){10}}
\end{picture} & $-0.06314$ & 

26 &
\begin{picture}(18,14)(0,5)
\put(5,1){\circle*{2}}
\put(15,1){\circle*{2}}
\put(15,11){\circle*{2}}
\put(5,11){\circle*{2}}
\put(5,1){\line(1,0){10}}
\put(5,11){\line(1,0){10}}
\put(5,1){\line(1,1){10}}
\end{picture} & $-0.29456$  & 

27 &
\begin{picture}(18,14)(0,5)
\put(5,1){\circle*{2}}
\put(15,1){\circle*{2}}
\put(15,11){\circle*{2}}
\put(5,11){\circle*{2}}
\put(5,0.5){\line(1,0){10}}
\put(5,1.5){\line(1,0){10}}
\put(5,11){\line(1,0){10}}
\put(5,1){\line(1,1){10}}
\end{picture} & $-0.04937$ \\ 
 & & & & & & & & \\

28 &
\begin{picture}(18,14)(0,5)
\put(5,1){\circle*{2}}
\put(15,1){\circle*{2}}
\put(15,11){\circle*{2}}
\put(5,11){\circle*{2}}
\put(5,1){\line(1,0){10}}
\put(5,11){\line(1,0){10}}
\put(5,0.5){\line(1,1){10}}
\put(5,1.5){\line(1,1){10}}
\end{picture} & $0.17930$  & 

29 &
\begin{picture}(18,14)(0,5)
\put(5,1){\circle*{2}}
\put(15,1){\circle*{2}}
\put(15,11){\circle*{2}}
\put(5,11){\circle*{2}}
\put(5,1){\line(1,0){10}}
\put(5,11){\line(1,0){10}}
\put(5,1){\line(0,1){10}}
\end{picture} & $-0.15733$  & 

30 &
\begin{picture}(18,14)(0,5)
\put(5,1){\circle*{2}}
\put(15,1){\circle*{2}}
\put(15,11){\circle*{2}}
\put(5,11){\circle*{2}}
\put(5,0.5){\line(1,0){10}}
\put(5,1.5){\line(1,0){10}}
\put(5,11){\line(1,0){10}}
\put(5,1){\line(0,1){10}}
\end{picture} & $-0.15941$ \\ 
 & & & & & & & & \\

31 &
\begin{picture}(18,14)(0,5)
\put(5,1){\circle*{2}}
\put(15,1){\circle*{2}}
\put(15,11){\circle*{2}}
\put(5,11){\circle*{2}}
\put(5,1){\line(1,0){10}}
\put(5,11){\line(1,0){10}}
\put(4.5,1){\line(0,1){10}}
\put(5.5,1){\line(0,1){10}}
\end{picture} & $0.11217$  & 

32 &
\begin{picture}(18,14)(0,5)
\put(5,1){\circle*{2}}
\put(15,1){\circle*{2}}
\put(15,11){\circle*{2}}
\put(5,11){\circle*{2}}
\put(5,1){\line(1,0){10}}
\put(5,11){\line(1,0){10}}
\put(5,1){\line(0,1){10}}
\put(15,1){\line(0,1){10}}
\end{picture} & $0.47978$ &
 & & \\
& & & & & & & & \\
\end{tabular}
\caption{Couplings used for the FP action.}
\label{couplings}
\end{table}

\begin{table} 
\setlength{\unitlength}{0.7mm}
\begin{tabular}{c c c c c c c c c}
No. & Type & Coeff. & No. & Type & Coeff.
& No. & Type & Coeff. \\
\hline

1 &
\begin{picture}(20,20)(0,4)
\put(0,0){\dashbox{1}(10,10)}
\put(0,10){\dashbox{1}(10,10)}
\put(10,0){\dashbox{1}(10,10)}
\put(10,10){\dashbox{1}(10,10)}
\put(7.5,7.5){\line(1,0){1.5}}
\put(7.5,7.5){\line(0,1){1.5}}
\put(7.5,7.5){\line(-1,0){1.5}}
\put(7.5,7.5){\line(0,-1){1.5}}
\put(9,9){\framebox(2,2)}
\end{picture} & 
$0.59497$
&

2 &
\begin{picture}(20,20)(0,4)
\put(0,0){\dashbox{1}(10,10)}
\put(0,10){\dashbox{1}(10,10)}
\put(10,0){\dashbox{1}(10,10)}
\put(10,10){\dashbox{1}(10,10)}
\put(7.5,7.5){\line(1,0){1.5}}
\put(7.5,7.5){\line(0,1){1.5}}
\put(7.5,7.5){\line(-1,0){1.5}}
\put(7.5,7.5){\line(0,-1){1.5}}
\put(-1,9){\framebox(2,2)}
\end{picture} & 
$0.15621$
&

3 &
\begin{picture}(20,20)(0,4)
\put(0,0){\dashbox{1}(10,10)}
\put(0,10){\dashbox{1}(10,10)}
\put(10,0){\dashbox{1}(10,10)}
\put(10,10){\dashbox{1}(10,10)}
\put(7.5,7.5){\line(1,0){1.5}}
\put(7.5,7.5){\line(0,1){1.5}}
\put(7.5,7.5){\line(-1,0){1.5}}
\put(7.5,7.5){\line(0,-1){1.5}}
\put(-1,-1){\framebox(2,2)}
\end{picture} &  
$0.08300$ \\
 & & & & & & & & \\

4 &
\begin{picture}(20,20)(0,4)
\put(0,0){\dashbox{1}(10,10)}
\put(0,10){\dashbox{1}(10,10)}
\put(10,0){\dashbox{1}(10,10)}
\put(10,10){\dashbox{1}(10,10)}
\put(7.5,7.5){\line(1,0){1.5}}
\put(7.5,7.5){\line(0,1){1.5}}
\put(7.5,7.5){\line(-1,0){1.5}}
\put(7.5,7.5){\line(0,-1){1.5}}
\put(-1,19){\framebox(2,2)}
\end{picture} & 
$0.00942$
&

5 &
\begin{picture}(20,20)(0,4)
\put(0,0){\dashbox{1}(10,10)}
\put(0,10){\dashbox{1}(10,10)}
\put(10,0){\dashbox{1}(10,10)}
\put(10,10){\dashbox{1}(10,10)}
\put(7.5,7.5){\line(1,0){1.5}}
\put(7.5,7.5){\line(0,1){1.5}}
\put(7.5,7.5){\line(-1,0){1.5}}
\put(7.5,7.5){\line(0,-1){1.5}}
\put(9,19){\framebox(2,2)}
\end{picture} & 
$-0.00171$
&

6 &
\begin{picture}(20,20)(0,4)
\put(0,0){\dashbox{1}(10,10)}
\put(0,10){\dashbox{1}(10,10)}
\put(10,0){\dashbox{1}(10,10)}
\put(10,10){\dashbox{1}(10,10)}
\put(7.5,7.5){\line(1,0){1.5}}
\put(7.5,7.5){\line(0,1){1.5}}
\put(7.5,7.5){\line(-1,0){1.5}}
\put(7.5,7.5){\line(0,-1){1.5}}
\put(19,19){\framebox(2,2)}
\end{picture} &  
$-0.00668$ \\
 & & & & & & & & \\

7 &
\begin{picture}(20,20)(0,4)
\put(0,0){\dashbox{1}(10,10)}
\put(0,10){\dashbox{1}(10,10)}
\put(10,0){\dashbox{1}(10,10)}
\put(10,10){\dashbox{1}(10,10)}
\put(7.5,7.5){\line(1,0){1.5}}
\put(7.5,7.5){\line(0,1){1.5}}
\put(7.5,7.5){\line(-1,0){1.5}}
\put(7.5,7.5){\line(0,-1){1.5}}
\put(9,9){\framebox(2,2)}
\put(10,10){\circle*{1.5}}
\put(0,10){\circle*{1.5}}
\put(0,10){\line(1,0){10}}
\end{picture} & 
$-0.03600$
&

8 &
\begin{picture}(20,20)(0,4)
\put(0,0){\dashbox{1}(10,10)}
\put(0,10){\dashbox{1}(10,10)}
\put(10,0){\dashbox{1}(10,10)}
\put(10,10){\dashbox{1}(10,10)}
\put(7.5,7.5){\line(1,0){1.5}}
\put(7.5,7.5){\line(0,1){1.5}}
\put(7.5,7.5){\line(-1,0){1.5}}
\put(7.5,7.5){\line(0,-1){1.5}}
\put(9,9){\framebox(2,2)}
\put(10,10){\circle*{1.5}}
\put(0,0){\circle*{1.5}}
\put(0,0){\line(1,1){10}}
\end{picture} & 
$-0.03104$
&

9 &
\begin{picture}(20,20)(0,4)
\put(0,0){\dashbox{1}(10,10)}
\put(0,10){\dashbox{1}(10,10)}
\put(10,0){\dashbox{1}(10,10)}
\put(10,10){\dashbox{1}(10,10)}
\put(7.5,7.5){\line(1,0){1.5}}
\put(7.5,7.5){\line(0,1){1.5}}
\put(7.5,7.5){\line(-1,0){1.5}}
\put(7.5,7.5){\line(0,-1){1.5}}
\put(9,9){\framebox(2,2)}
\put(10,0){\circle*{1.5}}
\put(0,10){\circle*{1.5}}
\put(0,10){\line(1,-1){10}}
\end{picture} & 
$-0.11141$ \\
 & & & & & & & & \\

10 &
\begin{picture}(20,20)(0,4)
\put(0,0){\dashbox{1}(10,10)}
\put(0,10){\dashbox{1}(10,10)}
\put(10,0){\dashbox{1}(10,10)}
\put(10,10){\dashbox{1}(10,10)}
\put(7.5,7.5){\line(1,0){1.5}}
\put(7.5,7.5){\line(0,1){1.5}}
\put(7.5,7.5){\line(-1,0){1.5}}
\put(7.5,7.5){\line(0,-1){1.5}}
\put(-1,9){\framebox(2,2)}
\put(10,10){\circle*{1.5}}
\put(0,10){\circle*{1.5}}
\put(0,10){\line(1,0){10}}
\end{picture} & 
$0.02788$
&

11 &
\begin{picture}(20,20)(0,4)
\put(0,0){\dashbox{1}(10,10)}
\put(0,10){\dashbox{1}(10,10)}
\put(10,0){\dashbox{1}(10,10)}
\put(10,10){\dashbox{1}(10,10)}
\put(7.5,7.5){\line(1,0){1.5}}
\put(7.5,7.5){\line(0,1){1.5}}
\put(7.5,7.5){\line(-1,0){1.5}}
\put(7.5,7.5){\line(0,-1){1.5}}
\put(-1,9){\framebox(2,2)}
\put(10,10){\circle*{1.5}}
\put(10,0){\circle*{1.5}}
\put(10,0){\line(0,1){10}}
\end{picture} & 
$-0.01814$ 
&

12 &
\begin{picture}(20,20)(0,4)
\put(0,0){\dashbox{1}(10,10)}
\put(0,10){\dashbox{1}(10,10)}
\put(10,0){\dashbox{1}(10,10)}
\put(10,10){\dashbox{1}(10,10)}
\put(7.5,7.5){\line(1,0){1.5}}
\put(7.5,7.5){\line(0,1){1.5}}
\put(7.5,7.5){\line(-1,0){1.5}}
\put(7.5,7.5){\line(0,-1){1.5}}
\put(-1,9){\framebox(2,2)}
\put(10,10){\circle*{1.5}}
\put(0,0){\circle*{1.5}}
\put(0,0){\line(1,1){10}}
\end{picture} & 
$-0.01660$ \\
 & & & & & & & & \\

13 &
\begin{picture}(20,20)(0,4)
\put(0,0){\dashbox{1}(10,10)}
\put(0,10){\dashbox{1}(10,10)}
\put(10,0){\dashbox{1}(10,10)}
\put(10,10){\dashbox{1}(10,10)}
\put(7.5,7.5){\line(1,0){1.5}}
\put(7.5,7.5){\line(0,1){1.5}}
\put(7.5,7.5){\line(-1,0){1.5}}
\put(7.5,7.5){\line(0,-1){1.5}}
\put(-1,9){\framebox(2,2)}
\put(0,10){\circle*{1.5}}
\put(0,0){\circle*{1.5}}
\put(0,0){\line(0,1){10}}
\end{picture} & 
$0.01600$ 
&

14 &
\begin{picture}(20,20)(0,4)
\put(0,0){\dashbox{1}(10,10)}
\put(0,10){\dashbox{1}(10,10)}
\put(10,0){\dashbox{1}(10,10)}
\put(10,10){\dashbox{1}(10,10)}
\put(7.5,7.5){\line(1,0){1.5}}
\put(7.5,7.5){\line(0,1){1.5}}
\put(7.5,7.5){\line(-1,0){1.5}}
\put(7.5,7.5){\line(0,-1){1.5}}
\put(-1,-1){\framebox(2,2)}
\put(0,10){\circle*{1.5}}
\put(10,10){\circle*{1.5}}
\put(0,10){\line(1,0){10}}
\end{picture} & 
$-0.01311$
&

15 &
\begin{picture}(20,20)(0,4)
\put(0,0){\dashbox{1}(10,10)}
\put(0,10){\dashbox{1}(10,10)}
\put(10,0){\dashbox{1}(10,10)}
\put(10,10){\dashbox{1}(10,10)}
\put(7.5,7.5){\line(1,0){1.5}}
\put(7.5,7.5){\line(0,1){1.5}}
\put(7.5,7.5){\line(-1,0){1.5}}
\put(7.5,7.5){\line(0,-1){1.5}}
\put(-1,-1){\framebox(2,2)}
\put(10,10){\circle*{1.5}}
\put(0,0){\circle*{1.5}}
\put(0,0){\line(1,1){10}}
\end{picture} & 
$0.03643$ \\
& & & & & & & & \\

16 &
\begin{picture}(20,20)(0,4)
\put(0,0){\dashbox{1}(10,10)}
\put(0,10){\dashbox{1}(10,10)}
\put(10,0){\dashbox{1}(10,10)}
\put(10,10){\dashbox{1}(10,10)}
\put(7.5,7.5){\line(1,0){1.5}}
\put(7.5,7.5){\line(0,1){1.5}}
\put(7.5,7.5){\line(-1,0){1.5}}
\put(7.5,7.5){\line(0,-1){1.5}}
\put(-1,-1){\framebox(2,2)}
\put(10,0){\circle*{1.5}}
\put(0,10){\circle*{1.5}}
\put(0,10){\line(1,-1){10}}
\end{picture} & 
$-0.02649$
&
 
17 &
\begin{picture}(20,20)(0,4)
\put(0,0){\dashbox{1}(10,10)}
\put(0,10){\dashbox{1}(10,10)}
\put(10,0){\dashbox{1}(10,10)}
\put(10,10){\dashbox{1}(10,10)}
\put(7.5,7.5){\line(1,0){1.5}}
\put(7.5,7.5){\line(0,1){1.5}}
\put(7.5,7.5){\line(-1,0){1.5}}
\put(7.5,7.5){\line(0,-1){1.5}}
\put(-1,-1){\framebox(2,2)}
\put(0,0){\circle*{1.5}}
\put(0,10){\circle*{1.5}}
\put(0,0){\line(0,1){10}}
\end{picture} & 
$0.01849$
&
 
& & \\
 & & & & & & & & \\

\end{tabular}
\caption{Coefficients of the parametrization of the fine field.}
\label{alphabeta}
\end{table}

\begin{table}\centering
\begin{tabular}{r@{} l r *{7}{r@{}l}}
\multicolumn{2}{c}{$N\beta$} & L & \multicolumn{2}{c}{$\xi$} &
\multicolumn{2}{c}{$\langle Q^2_{\rm coarse} \rangle $} & 
\multicolumn{2}{c}{$\langle Q^2_{\rm min\: 1.\, level} \rangle $} & 
\multicolumn{2}{c}{$\langle Q^2_{\rm par\: 1.\, level} \rangle $} &
\multicolumn{2}{c}{$\langle Q^2_{\rm par\: 2.\, level} \rangle $} & 
\multicolumn{2}{c}{$\chi^t_{\rm min\: 1.\, level} \xi^2$} &
\multicolumn{2}{c}{$\chi^t_{\rm par\: 1.\, level} \xi^2$} \\
\hline
1.&4   &   6 &  0.&9972(27) &  1.&0981(66) &  1.&008(17)   &    &        & %
                        &         & 0.&0278(5)  &   &          \\
2.&1   &  12 &  2.&0085(68) &  1.&8950(98) &  1.&758(22)   &   &        & %
                        &         & 0.&0492(7)  &   &          \\
2.&45  &  18 &  3.&044(14)  &  2.&270(13)  &  2.&157(28)   &  2.&121(26) & %
                      2.&114(44)  & 0.&0617(10) & 0.&0607(9)   \\
2.&45  &  34 &  2.&9500(71) &  8.&235(46)  &    &          &  7.&794(44) & %
                        &         &   &         & 0.&0587(4)   \\ 
2.&7   &  24 &  4.&267(19)  &  2.&272(16)  &  2.&152(31)   &    &        & %
                        &         & 0.&0680(12) &   &          \\
3.&0   &  36 &  6.&422(27)  &  2.&373(16)  &    &          &  2.&282(34) & %
                        &         &   &         & 0.&0726(12)  \\
3.&0   &  54 &  6.&241(23)  &  5.&477(39)  &    &          &  5.&208(37) & %
                        &         &   &         & 0.&0696(7)   \\
3.&0   &  76 &  6.&244(21)  & 11.&01(11)   &    &          & 10.&51(11)  & %
                        &         &   &         & 0.&0709(9)   \\
3.&2   &  48 &  8.&569(26)  &  2.&414(17)  &  2.&360(42)   &  2.&277(24) & %
                        &         & 0.&0752(14) & 0.&0726(9)   \\
3.&5   &  74 & 13.&063(71)  &  2.&410(31)  &    &          &  2.&304(29) & %
                        &         &   &         & 0.&0718(12)  \\
3.&5   &  94 & 12.&897(73)  &  3.&945(49)  &    &          &  3.&786(48) & %
                        &         &   &         & 0.&0713(12)  \\
3.&79  & 110 & 19.&945(83)  &  2.&217(23)  &    &          &  2.&138(36) & %
                        &         &   &         & 0.&0703(13)  \\
3.&79  & 200 & 19.&43(13)   &  7.&61(13)   &    &          &  7.&29(12)  & %
                        &         &   &         & 0.&0688(15)  \\
3.&95  & 150 & 24.&92(15)   &  2.&607(35)  &    &          &  2.&528(34) & %
                        &         &   &         & 0.&0698(13)  \\
\end{tabular}
\caption{Results of MC simulations.}
\label{mcres}
\end{table}

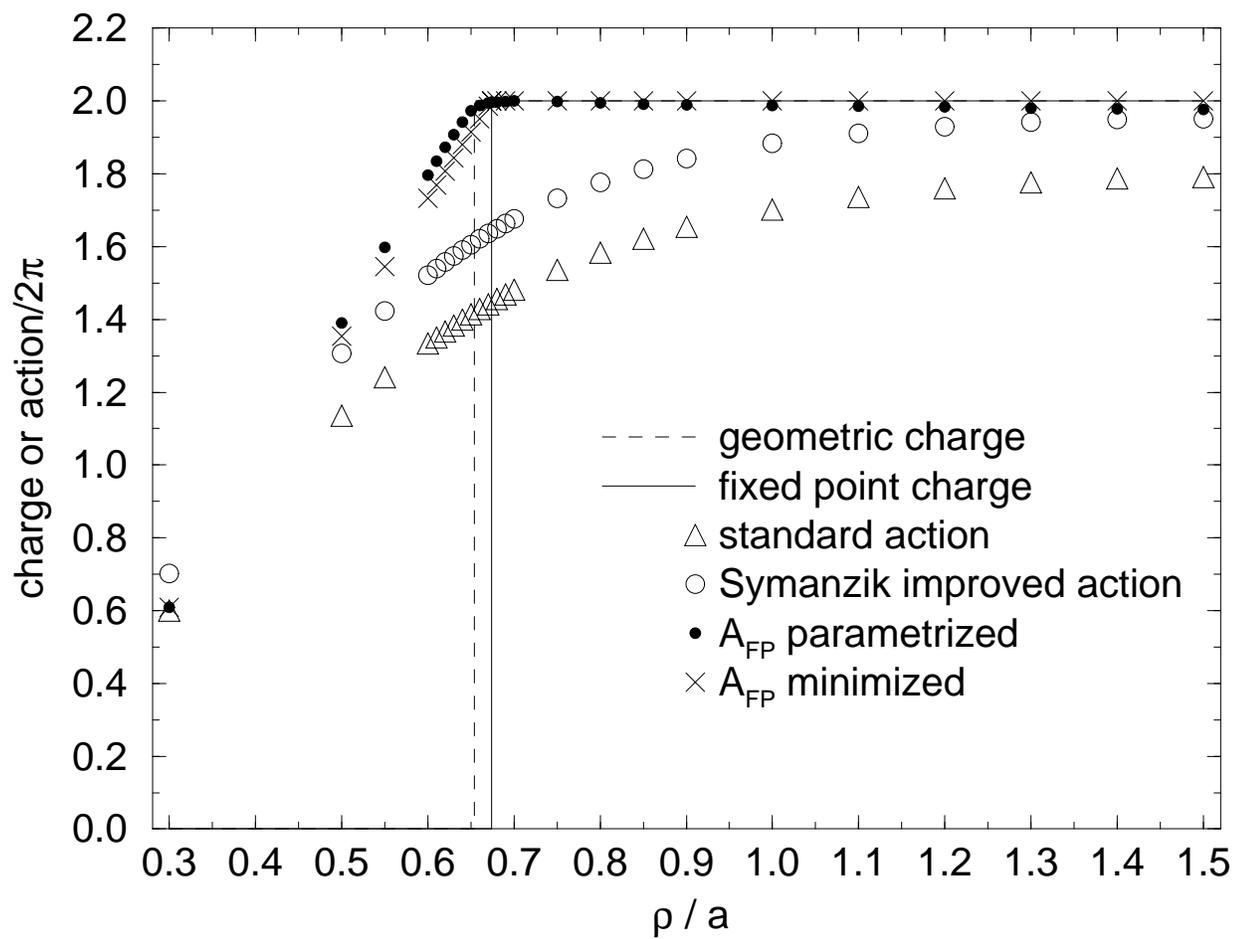
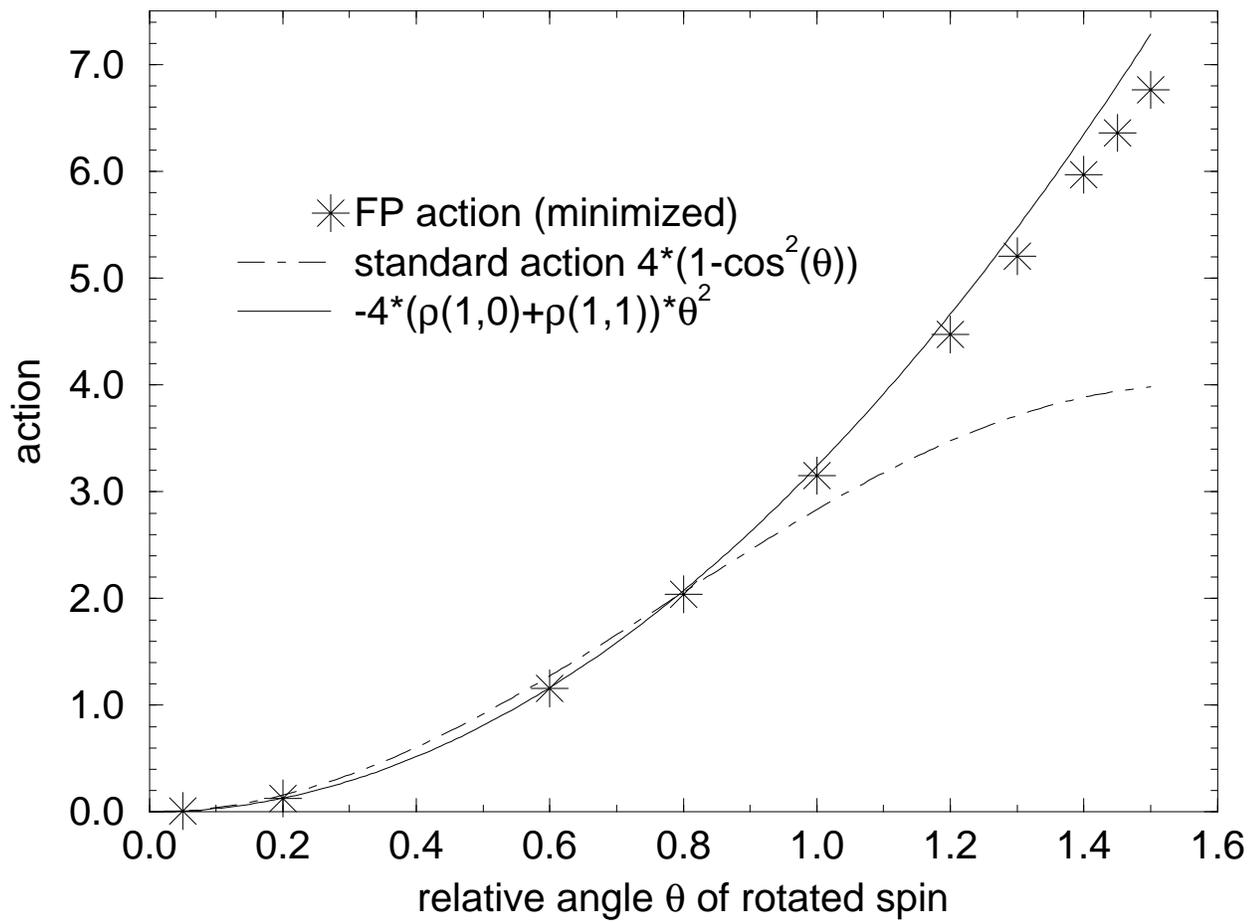
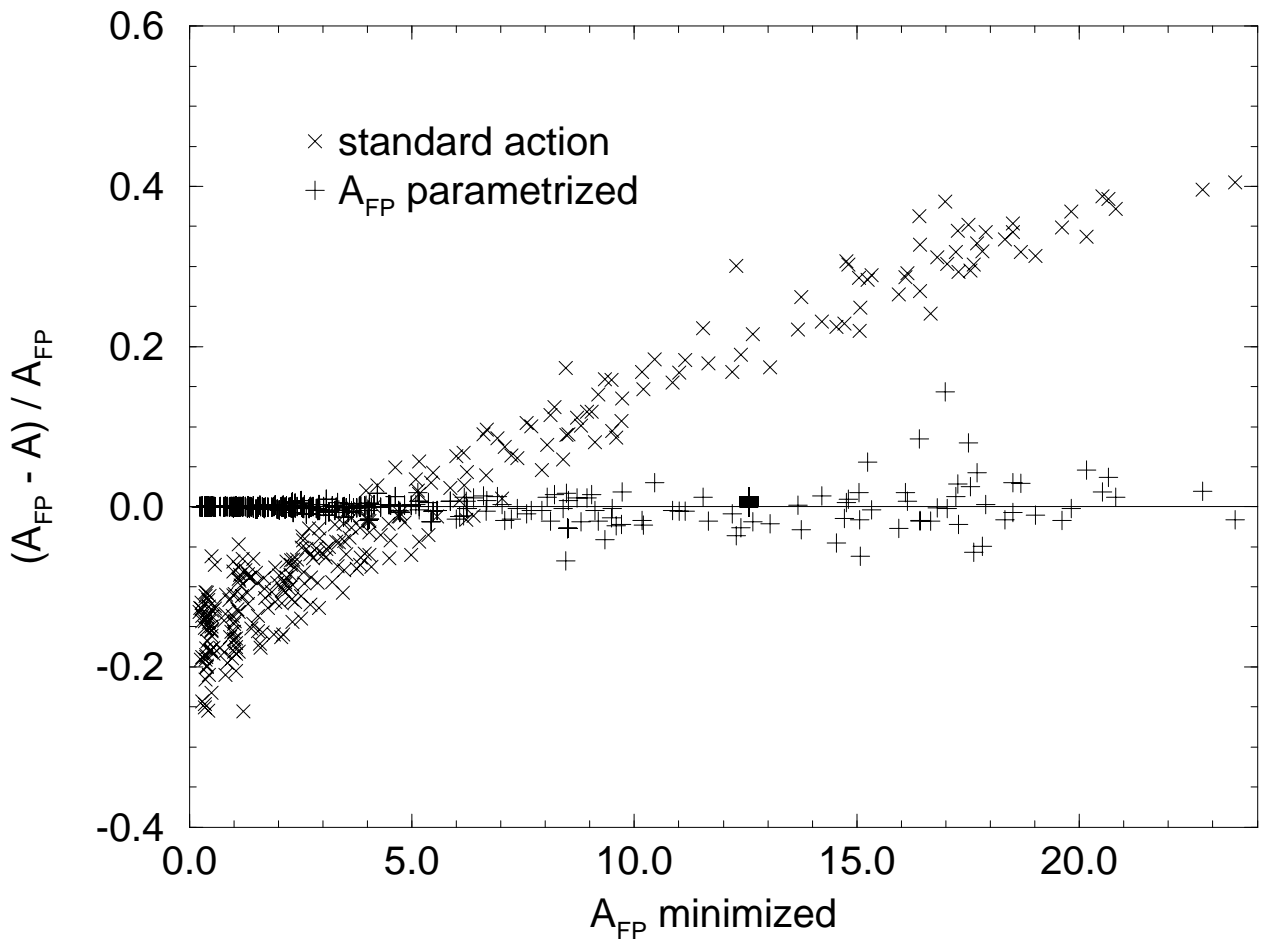
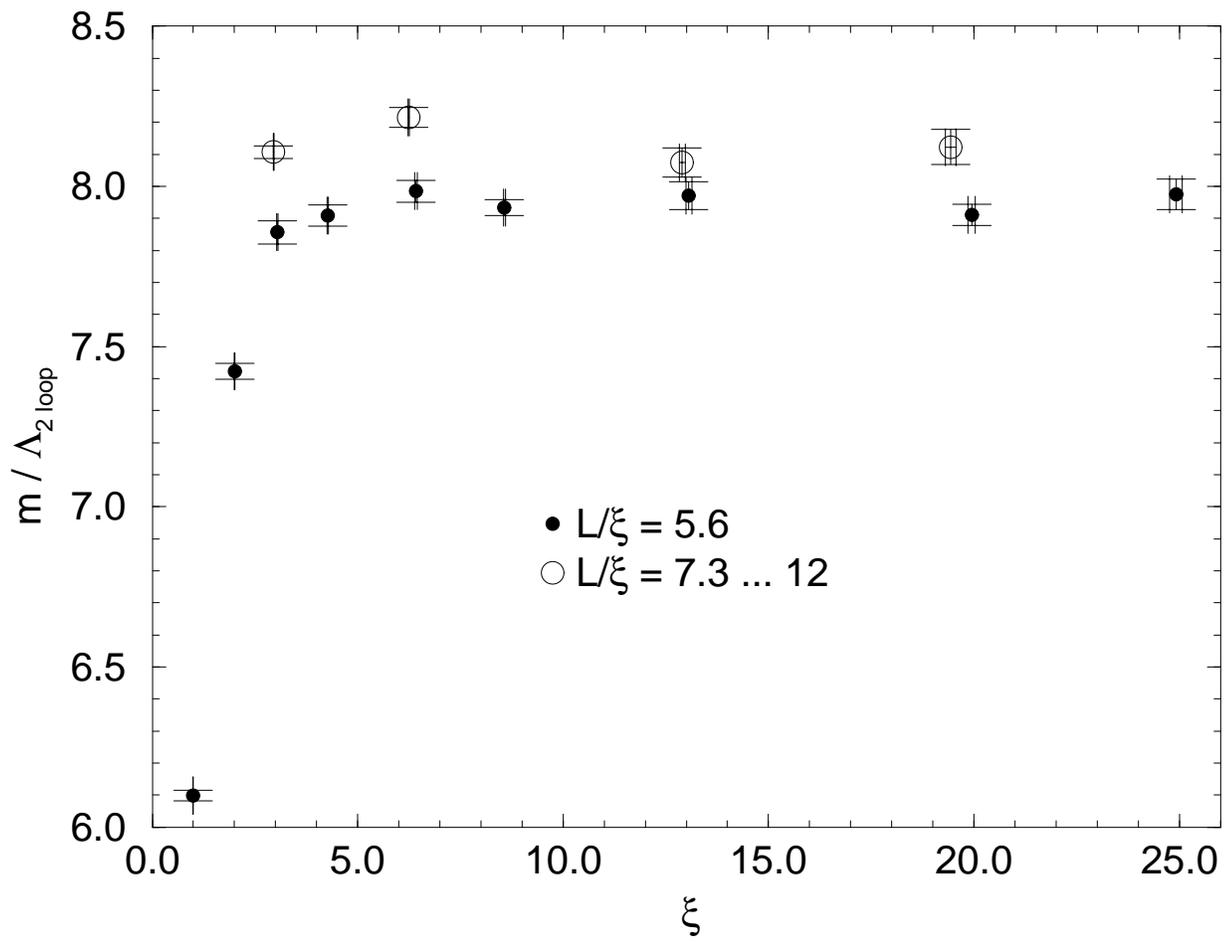
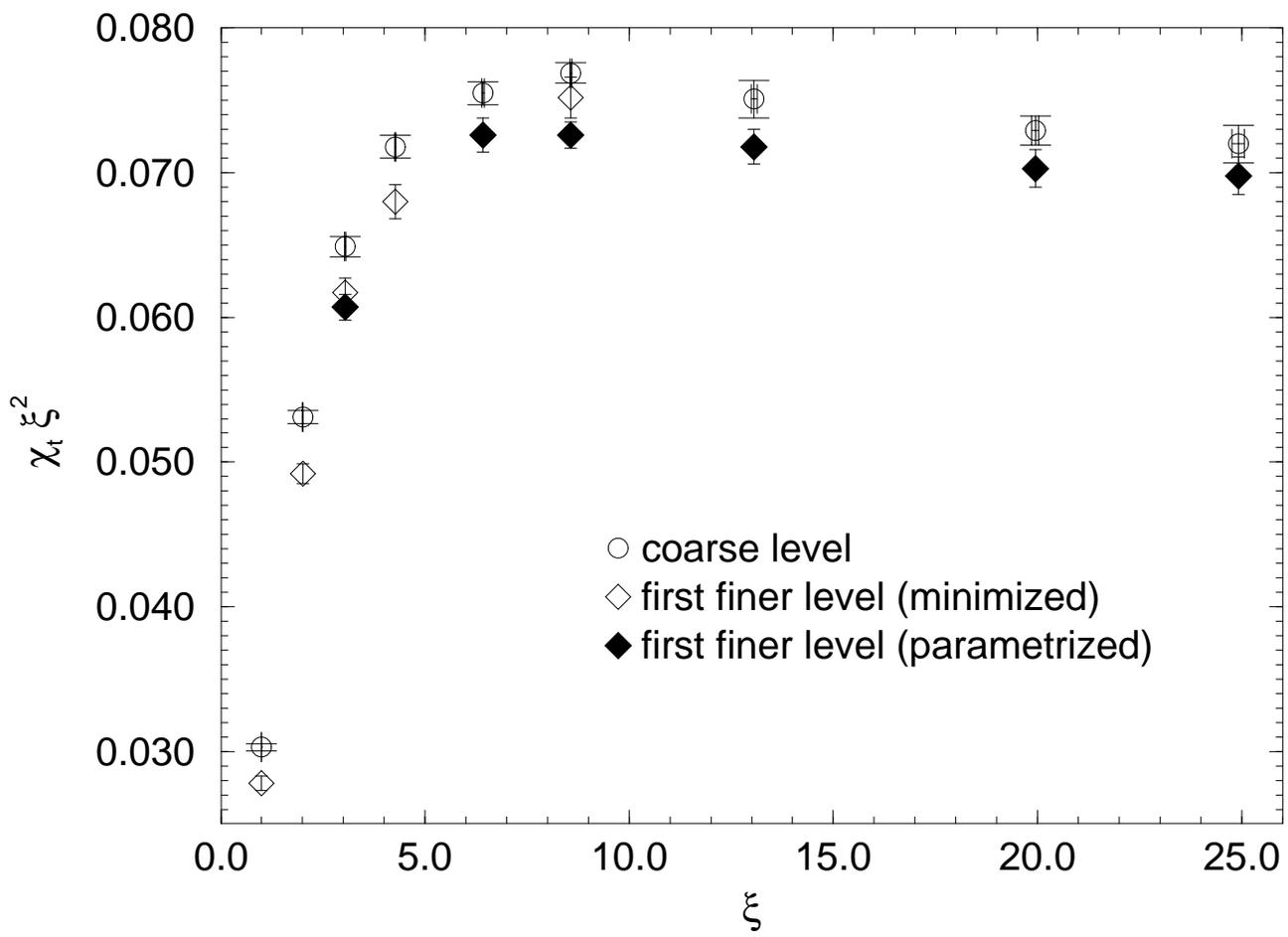
\end{document}